\documentclass[onecolum,secnumarabic,amssymb, nobibnotes,longbibliography,preprint]{revtex4-1}

\usepackage{graphicx,color,amsmath,hyperref,bm}
\usepackage[normalem]{ulem}

\begin{document}
	%\linenumbers
	
	\title{The two-thirds power law derived from an higher-derivative action}
	
	\author{N. Boulanger$^1$}
	\email[e-mail:]{nicolas.boulanger@umons.ac.be}
	\author{F. Buisseret$^{2,3}$}
	\email[e-mail:]{buisseretf@helha.be}
	\author{F. Dierick$^{4,5}$}
	\email[e-mail:]{frederic.dierick@gmail.com}
	\author{O. White$^6$}
	\email[e-mail:]{olivier.white@u-bourgogne.fr}
	
	\affiliation{$^1$ Service de Physique de l'Univers, Champs et Gravitation, Universit\'{e} de Mons, UMONS  Research Institute for Complex Systems, Place du Parc 20, 7000 Mons, Belgium }
	\affiliation{$^2$ CeREF, Chaussée de Binche 159, 7000 Mons, Belgium}
	\affiliation{$^3$ Service de Physique Nucl\'{e}aire et Subnucl\'{e}aire, Universit\'{e} de Mons, UMONS Research Institute for Complex Systems, 20 Place du Parc, 7000 Mons, Belgium}
		\affiliation{$^4$ Centre National de R\'{e}\'{e}ducation Fonctionnelle et de R\'{e}adaptation -- Rehazenter, RehaLAB, Luxembourg, Luxembourg}
		\affiliation{$^5$ Facult\'{e} des Sciences de la Motricit\'e, UCLouvain, 1-2 Place Pierre de Coubertin, 1348 Louvain-la-Neuve, Belgium}

	\affiliation{$^6$ Universit\'{e} de Bourgogne INSERM-U1093 Cognition, Action, and Sensorimotor Plasticity, Campus Universitaire, BP 27877, 21078 Dijon, France}

	\date{\today}%

\begin{abstract}
The two-thirds power law is a link between angular speed $\omega$ and  curvature $\kappa$ observed in voluntary human movements: $\omega$ is proportional to $\kappa^{2/3}$. Squared jerk is known to be a Lagrangian leading to the latter law. We propose that a broader class of higher-derivative Lagrangians leads to the two-thirds power law and we perform the Hamiltonian analysis leading to action-angle variables through Ostrogradski's procedure. In this framework, squared jerk appears as an action variable and its minimization may be related to power expenditure minimization during motion. The identified higher-derivative Lagrangians are therefore natural candidates for cost functions, i.e. movement functions that are targeted to be minimal when one individual performs a voluntary movement.
\end{abstract}

\maketitle

\section{Introduction}

Understanding the fundamental principles governing motions observed in Nature can be thought of as one of humanity's earliest philosophical endeavours: Making sense of the perceived physical world naturally leads to the study of mechanics. After the pioneering works of Newton (1643–1727), mechanics considerably evolved thanks to the reformulations of Lagrange (1736–1813) and  Hamilton (1805–1865), see e.g. \cite{L&L} for a detailed discussion of these formulations. 
The Newtonian approach relies on the fundamental equation $\vec F=m\, \vec a\,$ in Cartesian coordinates, where the total external force $\vec F$ acting on a pointlike body of inertial mass $m$ gives it an acceleration $\vec a=\ddot{\vec x}\,$, which in turn allows to yield the velocity $\vec v(t)=\dot{\vec x}$ and position $\vec x(t)$ of the body after successive integrations and upon imposing initial conditions. 
By invoking a principle of least action, Lagrange replaced Newton’s equation by the search for a single function named after him and denoted by $L\,$. In its original formulation, 
the Lagrangian $L$ depends on the configuration (position) space variables $q^\alpha(t)$ and their 
first-order time derivative $\dot{q}^\alpha(t)\,$, where $\alpha\in \{1,\ldots,n\}$ and the integer 
$n$ is the dimension of the configuration space, i.e., the number of degrees of freedom of the system 
once all the holonomic constraints have been taken into account.
Typically, $L=E_k - E_p\,$, the difference between the kinetic and potential energy functions of the system. 
Lagrange’s postulate is that the dynamical system, during its temporal evolution between two instants $t_i$ and $t_f$, 
will always minimize the action $S[q^\alpha]=\int^{t_f}_{t_i}L(q^\alpha,\dot{q}^\alpha)\, dt\,$. 
The  motion is then given by Hamilton's variational principle $\delta S=0$ subjected to the boundary conditions
$\delta q^\alpha(t_i)=0= \delta q^\alpha(t_f)\,$.

In the realm of motor control, that resides at the intersection of biomechanics, neuroscience, and mathematics, 
there has been a paradigm shift akin to the one from Newtonian to Lagrangian mechanics. Instead of merely describing motion, 
researchers have begun to ask why organisms move the way they do. Indeed, certain actions or behaviours are repeated in a consistent manner. The concept of stereotyped movements has been observed centuries ago, 
among others by Sherrington in 1906 \cite{Sherri}. The answer may lie in the minimization of a cost function, 
a mathematical representation of the effort, energy, or some other metric that a biological system tries to optimize. 
The interested reader may find reviews about optimal control theory in motor control in \cite{todorov02,todorov04}. 
By quantifying the cost linked with diverse trajectories or control strategies, these functions illuminate pathways to 
optimal movement strategies.

Traditionally applied to describe the dynamics of non-living physical systems, the recent introduction of Lagrangians as cost functions 
offers new insights accompanied by new challenges. One frequently cited cost functions in motor control relies on what 
is called the \emph{jerk}, $\vec j=\dot{\vec a}$, a vectorial quantity related to movement smoothness \cite{Bala15}. 
Research suggests that when humans make smooth, unperturbed movements, the trajectory they follow tends to minimize 
the averaged squared jerk, i.e., 
the movement minimizes the following action \cite{todorov98}
\begin{equation}\label{j0}
	S_J[\vec{x}(t)]=\int_{t_i}^{t_f} L_J\,dt =\int_{t_i}^{t_f} \lVert {\vec j} \rVert^2 \;dt\, .
\end{equation}
An apparent paradox that we aim to solve in the present study is that, from a mechanical point of view, 
the Lagrangian $L_J = \lVert {\vec j} \rVert^2$ does not lead to bounded trajectories. 
It is however considered in the study of bounded trajectories. We propose an other framework in which higher-derivative 
actions produce bounded trajectories satisfying two-thirds law, see below, 
and in which the minimization of jerk appears 
to be a dynamical consequence of the variational principle. 
By higher-derivative we mean a Lagrangian $L(\vec x^{(0)},\vec x^{(1)},\ldots , \vec x^{(N)})$ depending on 
the first $N>1$ derivatives of the dynamical variables $\vec x(t):=\vec x^{(0)}(t)$, i.e., depending 
on $\vec x^{(0)}(t)\,$, $\vec x^{(1)}(t):=\dot{ \vec x}(t)\,$, $\vec x^{(2)}(t):=\ddot{ \vec x}(t)\,$, etc. 
Three-dimensional vectors are assumed throughout. 
Study of such higher-derivative classical systems 
with finitely many degrees of freedom will be achieved by resorting  to Ostrogradski's approach  \cite{Ostrogradsky:1850fid}. 

Although cost functions can be central in elucidating optimal movement trajectories, they are not the only framework 
in motor control research. Another approach is to identify conserved quantities or invariants. 
Invariants, as the term implies, are quantities whose value stays constant during dynamical evolution. 
A particularly illuminating example of this field is the empirical two-thirds power law, hereafter referred to as 2/3-PL. 
It has been found in \cite{LACQUANITI1983115} and can be written as 
\begin{equation}\label{2_3_l}
	v=C\ \kappa^{-1/3},
\end{equation}
with velocity $\vec v=\dot{\vec x}$, speed $v=\lVert\vec v\rVert$ and $C\in\mathbb{R}^+_0$ a constant. 
One speaks of ``two-thirds" because this law's original formulation involved the angular speed 
$\omega=v\, \kappa$ where $\kappa=1/R$ is the curvature (the inverse of the radius $R$ of the osculating circle) 
of the trajectory at the given time, leading to $\omega =C\ \kappa^{2/3}$, i.e. the $2/3$ exponent. 
The initial observation was that the speed of a drawing or writing movement is related to the curvature of the drawing. 
This law has since been observed in a wide range of planar movements, see e.g. the review \cite{Zago2018}, 
and especially in elliptic trajectories. Note that a more general link of the type
\begin{equation}\label{2_3_l2}
	v=C\ \kappa^{\beta}
\end{equation}
may be observed in an even wider class of  trajectories \cite{huh2015}, though we will not consider this generalization here. 
There has been a lot of debate on 2/3-PL, some authors claiming that it is mostly an artefact due to fitting procedures, 
while others (including the authors of this paper) argue that this law is indeed a behavioural consequence of a fundamental law in human motion \cite{Zago2018}. 
Results from various studies challenge a purely kinematic interpretation of the 2/3-PL and highlight the role of the central nervous system 
in motion planning, that leads to (\ref{2_3_l}), using motor imagery paradigms for covert 
movements \cite{karlinsky14,Papa12}.

 This investigation aims to extend the understanding of cost functions through the lens of higher-derivative Lagrangians. It proposes that a class of higher-derivative actions broader than (\ref{j0}) leads to the 2/3-PL, offering new insights into cost functions in human motion. 
This is developed in Sec.~\ref{sec:lag} after general considerations about 2/3-PL in Sec.~\ref{sec:2_3}. 

Complementing the Lagrangian perspective, the Hamiltonian formalism provides a phase-space representation 
of the dynamics of a system as well as a general way to compute invariants through action-angle variables \cite{L&L}. 
The proposed higher-derivative Lagrangians will give rise to corresponding Hamiltonian functions through 
Ostrogradsky's procedure \cite{Ostrogradsky:1850fid}, after which the invariants will be computed in Sec. \ref{sec:ham}. 
A contextualisation of these results in the framework of motor control will then be discussed in Sec.~\ref{sec:discu}.

\section{Two-thirds law: Kinematical considerations}\label{sec:2_3}
In the context of three-dimensional motion in Euclidean space, the curvature and torsion of a given trajectory are 
obtained through the well-known formulae
\begin{equation}\label{T_K}
	\kappa=\frac{\lVert \vec v \times \vec a \rVert}{v^3}\;, \quad 
	T = \frac{\vec j\cdot(\vec v\times \vec a)}{\lVert\vec v\times \vec a\rVert^2}\;,
\end{equation}
where $\vec a=\ddot{\vec x}$ and where the symbol $\times$ stands for the usual vector product in three-dimensional 
space. 
From the above definitions, one can write
\begin{equation}
	v=\lVert\vec v\times \vec a\rVert^{1/3}\, \kappa^{-1/3}\;.
\end{equation}
Hence, 2/3-PL is valid if the norm of the vector $\,\vec \ell_2=\vec v\times\vec a\,$ is constant. 
A sufficient condition is $\dot{\vec{\ell}}_2=\vec 0$, implying that 
\begin{equation}\label{cond}
	\vec j=\gamma\, \vec v\;. 
\end{equation}
The coefficient function $\gamma$ may explicitly depend on time and on the various derivatives $\vec x^{\, (i)}$. 
Therefore, a trajectory such that 
$\vec x^{(3)} - \gamma(\vec x^{(i)},t)\, \vec x^{\, (1)}=\vec 0$ 
satisfies (\ref{2_3_l}). The simplest choice is that of a constant function $\gamma$, leading to elliptic trajectories. 
In \cite{MATIC2020102453}, different choices of the form $\gamma=\gamma(t)$ are explored, 
leading to trajectories that all comply with 2/3-PL.

Condition (\ref{cond}) implies that the motion is planar, so that the torsion vanishes, $T=0$. 
Hence, non planar trajectories should not be related to (\ref{2_3_l}). Although it is not the main topic of our paper, it has been proposed in \cite{POLLICK2009325} that the law
\begin{equation}\label{3Dlaw}
	v= C \, \kappa^{-1/3}\, T^{-1/6}
\end{equation}
should hold in non-planar, three-dimensional motion, with $C$ a positive constant. 
From (\ref{T_K}) it can be deduced that $C=\lVert\vec j\cdot(\vec v\times\vec a)\rVert^{1/6}$. 
In other words, the law is valid if $\vec{j}\cdot(\vec v\times\vec a)$ is constant, or 
\begin{equation}
	\dot{\vec j} = f\,\vec v \;.
\end{equation}
In the spirit of \cite{MATIC2020102453}, one can say that any trajectory solution of 
\begin{equation}\label{3Ded}
\vec x^{\, (4)}-f(\vec x^{\, (i)},t)\, \vec x^{\, (1)}=\vec 0
\end{equation} 
will satisfy (\ref{3Dlaw}).  

\section{Dynamical principle -- Lagrangian formalism}\label{sec:lag}

\subsection{The model}
We propose that the  actions associated with the higher derivative Lagrangians
\begin{equation}\label{L0first}
	L=\frac{\lambda}{2} \,\lVert \vec x^{\, (N)}\rVert ^2-\frac{1}{2}\,U\left(\lVert \vec x^{\, (N-1)}\rVert ^2\right),
\end{equation}
with  $N\geq 1$ and $\lambda\in\mathbb{R}^+_0$, are relevant cost function candidates that may lead to trajectories satisfying (\ref{2_3_l}) for any ``potential" function $U(z)\,$ where the variable $z$ denotes $\lVert \vec x^{\, (N-1)}\rVert ^2\,$. 
Note that the above Lagrangian is higher-derivative as soon as $N>1$.

From the variational principle based on the action functional $S[\vec{x}(t)]=\int_{t_i}^{t_f} L\,dt$, 
the equations of motion read as follows, for a generic higher-derivative Lagrangian:
\begin{equation}
\vec	0 = \frac{\delta L}{\delta \vec x} \equiv \sum^N_{j=0} \left(-\frac{d}{dt}\right)^j\, 
	\frac{\partial L}{\partial \vec x^{(j)}} \;.
	\label{eom}
\end{equation}
These equations have to be satisfied together with the vanishing of the boundary terms 
defining the momenta $\vec p_i\,$:
\begin{eqnarray}
	\left. \sum_{i=0}^{N-1} \delta \vec x^{(i)}\, \vec p_{i}\right|^{t_2}_{t_1}\;\; &=& 0\;,
	\nonumber \\
	\vec p_{i}&:=& \frac{\delta L}{\delta \vec x^{(i+1)}}
	\equiv \sum^{N-i-1}_{j=0} \left(-\frac{d}{dt}\right)^j\, 
	\frac{\partial L}{\partial\vec  x^{(i+j+1)}}
	\;, \quad i\in\{0,\ldots,N-1\}\;.
	\label{momenta}
\end{eqnarray}
One chooses to cancel the above boundary terms by imposing the following conditions at the boundaries of the integration domain:
\begin{equation}
	\delta \vec x^{(j)}(t_f) = 0 = \delta \vec x^{(j)}(t_i)\;, \quad \forall ~j\in\{0,\ldots,N-1\}\;. \nonumber
\end{equation}
Since our Lagrangians do not explicitly depend on time, we will choose the initial date $t_i = 0$ for convenience. 
More specifically, the equations of motion computed from the Lagrangian (\ref{L0first})  are
\begin{equation}
\label{EOMa}
\lambda\ \frac{d^N}{dt^N}\,\vec x^{\, (N)} = - \frac{d^{N-1}}{dt^{N-1}} \left[U'(\lVert \vec x^{(N-1)}\rVert^2)\ \vec x^{\, (N-1)}\right]\; .
\end{equation}
Integrating them $N-1$ times leads to
\begin{equation}
\label{integratingEOM}
	\lambda\ \vec x^{\, (N+1)}=-U'(\vec x^{\, (N-1)\, 2})\ \vec x^{\, (N-1)}
	 + \sum^{N-2}_{j=0} \frac{\vec b_j}{j!} \, t^j \;,
\end{equation}   
in terms of $N-1$ constant vectors $\{\vec b_j\}\,$, $j = 0, \ldots, N-2\,$ that can be fixed by initial conditions.
As we see from the right-hand side of the above equation, the Lagrangian field equations \eqref{EOMa} 
lead to general solutions with a polynomial dependence on the evolution parameter, 
therefore signalling an instability, that is a landmark of higher-derivative models. 

We see that in order to avoid any instabilities -- also called \emph{run-away solutions} in the context 
of field theory --, we have to impose the initial conditions 
\begin{equation}
\label{initialcond}
\vec b_j=\vec 0\,, \quad \forall\;j \in \{0, \ldots, N-2\}\;,
\end{equation}
because then, the field equations \eqref{EOMa} lead to the differential equation  
\begin{equation}\label{diffequ}
\lambda\ \frac{d}{dt}\,\vec x^{(N)}+U'(\vec x^{\, (N-1)\, 2})\ \vec x^{\, (N-1)}=0\;
\end{equation}
that admits stable solutions and may imply the 2/3-PL in cases discussed below. 
Note that (\ref{initialcond}) only fixes $N-1$ conditions among the $2N$ initial conditions needed to ensure a unique solution.

An equivalent way of presenting the initial conditions \eqref{initialcond} is by first defining the $N-1$ vectors
\begin{equation}
 \vec{A}^N_j := \lambda\ \frac{d^{j+1}}{dt^{j+1}}\,\vec x^{(N)} + \frac{d^{j}}{dt^{j}}\, [ U'(\lVert \vec x^{(N-1)}\rVert^2)\, \vec x^{\, (N-1)}]\;,
 \qquad j \in \{0,\ldots,N-2\} \;,
\end{equation}
and then setting the initial conditions
\begin{equation}\label{initial1}
	\left.	\vec{A}^N_j \right|_{t=0} \equiv \vec b_j = \vec 0 \;,\quad {\forall}~ j\in \{0,\dots,N-2\}\;.
\end{equation}
The point with the latter presentation of the initial conditions \eqref{initialcond} is that, for the 
considered Lagrangian (\ref{L0first}), the set of vectors $\{\vec{A}^N_j\}\,$, $ j \in \{0,\ldots,N-2\} \,$ 
is in one-to-one correspondance with the set of momenta $\{\vec{p}_j\}\,$, $ j \in \{0,\ldots,N-2\} \,$, see  \eqref{momenta}.
In fact, one readily sees from Lagrangian (\ref{L0first}) and the definition \eqref{momenta} of the momenta that
\begin{equation}
\vec{A}^N_j = (-1)^{j-1}\,\vec{p}_{N-2-j}\;, \qquad j=0,\ldots, N-2\;.
\end{equation}
Therefore, the choice of initial conditions \eqref{initialcond}, equivalently \eqref{initial1}, is in turn equivalent to 
setting the following initial conditions on the first $N-1$ momenta: 
\begin{equation}\label{initialcondmomenta}
 \vec{p}_i(t=0) = \vec{0}\;,\qquad i = 0, 1, \ldots, N-2\;.
\end{equation}
To sum up the discussion, with the Lagrangian (\ref{L0first}) and the above initial conditions \label{initialcondmomenta}, 
the Lagrangian field equations \eqref{EOMa} lead to the differential equation
\begin{equation}
\label{EOM0}
\lambda\ \frac{d}{dt}\,\vec x^{(N)}+U'(\lVert \vec x^{(N-1)}\rVert^2)\ \vec x^{\, (N-1)}=0\;.
\end{equation}
From this equation, one observes that the vector 
\begin{equation}
\label{ell}
	\vec \ell_N=\lambda \,\vec{x}^{\, (N-1)}\times \vec x^{\, (N)}
\end{equation}
remains constant during the dynamical evolution, i.e., $\dot{\vec \ell}_N=\vec 0\,$.
Conversely, from the variational equations \eqref{EOMa} and its consequence \eqref{integratingEOM}, 
we see that imposing the condition $\dot{\vec \ell}_N=\vec 0\,$ is equivalent to imposing the initial 
conditions  \eqref{initial1}. Therefore, a compact way of imposing the initial conditions \eqref{initial1}
is by imposing that $\dot{\vec \ell}_N=\vec 0\,$.

The question of finding a Lagrangian whose equations of motion are (\ref{3Ded}) arises naturally but, 
at this stage, it remains an open problem since we have not found any Lagrangian leading to (\ref{3Ded}). 
In the following, we will choose not to investigate further the case of non-planar trajectories.

\subsection{The case $N=2$}

Equations (\ref{EOM0}) and (\ref{cond}) obviously coincide if $N=2$. 
In this case, the Lagrangian (\ref{L0first}) reduces to the Flash-Handzel Lagrangian \cite{Flash2007}, i.e.,
\begin{equation}\label{LFH}
	L=\frac{\lambda}{2} \, \lVert\vec a\rVert^{2}-\frac{1}{2}\,U\left(\lVert \vec v\rVert^2\right).
\end{equation}
If one imposes the initial condition $\vec b_0 := \lambda\,\vec{j}(0)+ \vec{v}(0) \,U'(v^2(0))=0\,$,  
the vector $\vec \ell_2=\Vec v\times\vec a$ is conserved since (\ref{EOM0}) now reads 
    \begin{equation}\label{EOM0b}
    	\lambda\ \vec j=-U'(\vec v^{\, 2})\ \vec v\;.
    \end{equation}
The trajectories generated by any choice of $U$ will therefore satisfy 2/3-PL.

\subsection{The case of a linear potential fonction}\label{sec:PU}
Keeping $N$ arbitrary but setting $U(z)=\lambda\, \omega^2\, z$ leads to Pais-Uhlenbeck 
oscillators \cite{Pais:1950za} since (\ref{L0first}) now reads

\begin{equation}\label{L0}
	L_{PU}=\frac{\lambda}{2} \,\left(\vec x^{\, (N)\, 2}-\omega^2\, \vec x^{\, (N-1)\, 2}\right)\;.
\end{equation}

After $2N-2$ integrations, the equations of motions (\ref{EOMa}) reduce to 
%\textbf{[Nicolas: I inserted the factorials at the denominator]}
\begin{equation}
	\ddot{\vec x}=-\omega^2\vec x + \sum^{2N-3}_{j=0}\;,
	\frac{\vec b_j}{j!} \,t^j\;
\end{equation}
whose general solution is
\begin{equation}\label{PUx}
	\vec x(t)=\vec K_1\, \cos(\omega t)+\vec K_2\, \sin(\omega t)+\sum^{2N-3}_{j=0}\vec a_j\, \frac{t^j}{j!}
\end{equation}
with $\vec a_{2N-3}=\frac{\vec b_{2N-3}}{\omega^2}\,$,  $\vec a_{2N-4}=\frac{\vec b_{2N-4}}{\omega^2}\,$, 
the other $\vec a_j$ with $j\leqslant 2N-5$ being recursively given by
	\begin{equation}\label{recur1}
		\omega^2 \vec a_j + \vec a_{j+2} = \vec b_j\;,\quad j \in \{0, 1, \ldots, 2N-5\}\; .
	\end{equation}
Imposing the initial conditions $\vec b_j=\vec 0$ for all $j$ 
clearly implies that $\vec a_{j}=\vec 0$ for all $j\,$
and the dynamics reduces to elliptic trajectories $\vec x(t)=\vec K_1\, \cos(\omega t)+\vec K_2\, \sin(\omega t)\,$ without any run-away modes.
Indeed, with these initial conditions the equations of motion reduce to  
$\ddot{\vec x}+\omega^2\vec x=0\,$, hence $\vec \ell_2=\omega^2 \vec \ell_1$ and the angular momentum $\vec \ell_1$ 
is conserved, as in standard Newtonian mechanics with central forces. 

\subsection{The case of a vanishing potential function}

In the case where $U(z)=0$ one recovers the mean squared derivative cost functions of~\cite{Richardson8201}
\begin{equation}
  L_{MS}=\frac{\lambda}{2}\ \vec x^{\, (N)\, 2}\;, 
\end{equation}
whose equations of motion are $\vec x^{\, (2N)}=\vec 0$, leading to
%\textbf{[Nicolas : I added the factorial at the denominator]}  
\begin{equation}
\vec x(t)=\sum^{2N-1}_{j=0}\frac{\vec b_j}{j!} \,t^j\;. 
\end{equation}
Such trajectories are unbounded and do not satisfy (\ref{cond}). Still, Lagrangian of this type have been shown to successfully model pointing tasks since the seminal work \cite{Hogan}, using $N=3$.

\section{Hamiltonian formalism}\label{sec:ham}

\subsection{Ostrogradski's approach}

Following \cite{Ostrogradsky:1850fid}, the position degrees of freedom are defined as $\vec q^{\, j}=\vec x^{\, (j)}$ and the momenta $\vec p_j$ are defined by (\ref{momenta}), for $j=0,\dots,N-1$. Specifically for the Lagrangian given in (\ref{L0first}), 
\begin{eqnarray}
	\vec p_j&=&(-)^{N-j-1}\lambda \vec x^{\, 2N-j-1}+(-)^{N-j-1}\frac{d^{N-j-2}}{dt^{N-j-2}}\left(U'(\vec x^{\, (N-1)\, 2})\ \vec x^{\, (N-1)}\right) \label{momentum}\\
	&=&(-)^{N-j-1}\frac{d^{N-j-2}}{dt^{N-j-2}}\vec A^N_{0}\equiv (-)^{N-j-1}\,\vec A^N_{N-j-2}\;.
\end{eqnarray}

Then, the Hamiltonian reads
\begin{equation}
	H=\sum^{N-2}_{j=0}\vec p_j\cdot\vec q^{\, j+1}+\frac{\vec p_{N-1}^{\, 2}}{2\lambda}+\frac{1}{2}U\left((\vec q^{\, N-1})^{\, 2}\right)
\end{equation}
and Hamilton's equations $\dot{\vec q}^{\, j}=\frac{\partial H}{\partial \vec p_{j}}\,$, $\dot{\vec p}_{j}=-\frac{\partial H}{\partial \vec q^{\, j}}\,$ 
lead to
\begin{eqnarray}\label{HamEq}
	\dot{\vec q}^{\; 0}&=&\vec q^{\, 1}\;,\\
	&\vdots& \nonumber\\
	\dot{\vec q}^{\, N-2}&=&\vec q^{\, N-1}\;,  \\
	\dot{\vec q}^{\, N-1}&=&\frac{\vec{p}_{N-1}}{\lambda}\;, \label{qdotNmoins1}\\
	\dot{\vec p}_0&=&\vec 0\;,\\
		\dot{\vec p}_1&=&-\vec p_0\;,\\
	&\vdots&\nonumber\\
	\dot{\vec p}_{N-2}&=&-\vec p_{N-3}\;,  \\
	\dot{\vec p}_{N-1}&=&-\vec p_{N-2}-U'\left((\vec q^{\, N-1})^{\, 2}\right) \ \vec q^{\, N-1}\;.
	\label{lastHE}
\end{eqnarray}

On the phase-space hypersurface given by 
\begin{equation}\label{initial2}
\vec p_0=\dots=\vec p_{N-2}=\vec 0\;,
\end{equation}
one is led to  the Hamiltonian 
\begin{equation}\label{Ham0}
	\tilde H=\frac{p_{N-1}^{\, 2}}{2\lambda}+\frac{\ell_N^2}{2\lambda(q^{\, N-1})^{\, 2}}+\frac{1}{2}U\left((q^{\, N-1})^{\, 2}\right)
\end{equation}
with $\ell_N$ being the constant norm of (\ref{ell}) and where, from then on, 
we use the notation $V^2=\lVert \vec{V} \rVert^2$ for vectors. 
We recall that the conditions (\ref{initial2}) are equivalent 
to (\ref{initial1}). 

The trajectory in the plane $(q^{N-1},p_{N-1})$ is such that $\tilde H={\cal E}$ constant -- we use the notation ${\cal E}$ although 
the Hamiltonian does not a priori possess the dimension of an energy. For example, in the Pais-Uhlenbeck case discussed in Sec. \ref{sec:PU}, the trajectory is a closed loop whose equation is given by
\begin{equation}
	p^2_{N-1}+\frac{\ell^2_N}{(q^{\, N-1})^{\, 2}}+\lambda^2\omega^2\, (q^{N-1})^2=2\,\lambda\,{\cal E}
\end{equation}
for $q^{N-1}>0\,$.

A stability analysis of trajectories can be carried out by applying the method of \cite[Chapter 7]{jose:1998} to the 
Hamiltonian equations (\ref{HamEq})--\eqref{lastHE}. 
The latter equations can be written under a matrix form $\dot{\bm\zeta}=A\,\bm\zeta\,$, where $\bm{\zeta}$ 
is a vector containing the $2N$ coordinates in phase-space and where $A$ is a $2N\times 2N$ matrix. The eigenvalues of $A$ are 0 and $\pm \sqrt{U'(0)/\lambda}$. The zero mode is a global translation mode. If $U'(0)<0$ as in the harmonic oscillator case, the nonzero eigenvalues are complex conjugated and all trajectories are bounded, global translation excepted. If $U'(0)>0$, there necessarily exist unbounded trajectories even if the global translation mode is set to zero. The existence of unbounded trajectories preserving 2/3-PL is a prediction that could be experimentally studied, but that is out of the scope of this paper. 

\subsection{Action variables}

We now restrict our discussion to potentials $U$ for which there exists some values of ${\cal E}$ such that ${\cal E}=\frac{\ell_N^2}{2\lambda(q^{\, N-1})^{\, 2}}+\frac{1}{2}U\left((q^{\, N-1})^{\, 2}\right)$ has two finite, distinct, solutions, leading to bounded trajectories, that will appear as closed loops $\Gamma$ in the plane $(q^{N-1},p_{N-1})$
where we recall the notation $q^{N-1}=\lVert \vec{q}^{N-1}\rVert$, idem for $p_{N-1}\,$.
In this case it is known that the action variable $I_N=\frac{1}{2\pi}\oint_\Gamma p_{N-1}dq^{N-1}$ is a constant of motion \cite{L&L}.
It can be written in various ways, where one always use to parametrize the closed curve $\Gamma$ by 
the evolution parameter $t\,$. First, by using  \eqref{qdotNmoins1}, 
\begin{equation}
I_N = \frac{1}{2\pi}\,\int_{t_i}^{t_f} p_{N-1}\,\dot{q}^{N-1} dt = \frac{1}{2\pi\lambda}\,\int_{t_i}^{t_f} p_{N-1}^2 dt = 
\frac{\lambda}{2\pi}\,\int_{t_i}^{t_f} q^{N\,2} dt\;, 
\end{equation}
where $t_f-t_i$ is the movement period.
By starting from the equivalent expression $I_N=-\frac{1}{2\pi}\oint_\Gamma q^{N-1}\,dp_{N-1}$ and 
using the solution \eqref{diffequ}, equivalently  \eqref{qdotNmoins1} with \eqref{lastHE}, we obtain
\begin{equation}
I_N = \frac{\lambda^2}{2\pi}\int^{t_f}_{t_i}\frac{q^{\, (N+1)\, 2}}{U'(q^{\, (N-1)\, 2})}dt\;.
\end{equation}
For potential functions $U\,$ such that $U'(0)\neq 0\,$, we define 
\begin{equation}
	I^a_N=\frac{\lambda^2}{2\pi U'(0)}\int^{t_f}_{t_i} q^{\, (N+1)\, 2}dt
\end{equation}
for future convenience.

In the Pais-Uhlenbeck case, $I_N=I^a_N=\frac{\lambda}{2\pi \omega^2}\int^{t_f}_{t_i} q^{\, (N+1)\, 2}dt$ and $\tilde H=\omega(2I_N+\ell_N )$ \cite[Chapter 6]{jose:1998}. We also refer the reader to \cite{Boulanger:2018tue} for a discussion of action variables in the generic Pais-Uhlenbeck oscillator model.
It can be expected that  $I_N\approx I^a_N$ from the Taylor expansion $U(z)\approx U(0)+U'(0) z+\dots$ 
It is worth pointing out that the mean squared derivative cost functions used in \cite{Richardson8201} are equal to $I_N^a$. 

\section{Conclusion: first principles shaping voluntary motion}\label{sec:discu}

\begin{figure}
	\includegraphics[width=6cm]{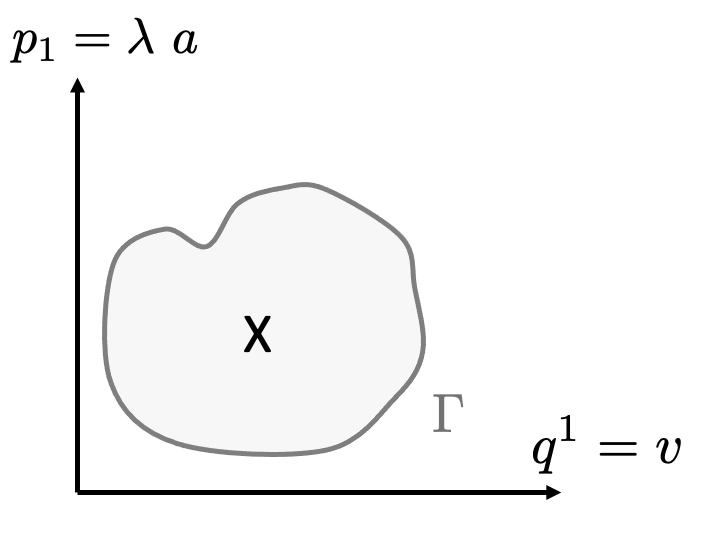}
	\caption{Typical phase-space plot of a bounded trajectory $\Gamma$ for $N=2$. The action variable $I_2$ is the area of $\Gamma$. The cross marks the average $q^1$ and $p_1$ on the loop $\Gamma$. }\label{fig}
\end{figure}

The application of Lagrangian and Hamiltonian formalisms in motor control demands a multidisciplinary approach that respects both a well-established mathematical formalism  and the intricacies of human physiology. This paper proposed a broader class of higher-derivative Lagrangians that, upon defining appropriate initial conditions, lead to trajectories complying with the 2/3-PL, thus providing new insights into cost functions critical to human motion. These Lagrangians are given in Eq.~(\ref{L0first}). A salient issue for the observation of 2/3-PL is the necessity of setting accurate initial conditions: If a Lagrangian involving up to the $N^{{\rm th}}$ time derivative $\vec x^{\, (N)}$ is used, it is necessary, in order 
to have a bounded motion, to set the Ostrogradski momenta $\vec p_{j}(0)=\vec 0$, with $0\leq j \leq N-2$. 
We recall that, for the class of Lagrangians \eqref{L0first}, the momenta are given by Eq.~(\ref{momentum}). 
If these $N-1$ initial conditions are unaligned with the natural capabilities of the human motor system, the considered Lagrangian is not qualified to model voluntary human movement. This consideration leads us to the conclusion that a minimal $N$ is the most natural choice. Therefore we think that $N=2$ actions of the form
\begin{equation}\label{LFH2}
	S=\int^{t_f}_{t_i}\left[\frac{\lambda}{2} \vec a^{\, 2}-\frac{1}{2}U\left(\vec v^{\, 2}\right)\right]dt
\end{equation}
are favoured:
\begin{itemize}
	\item They naturally lead to 2/3-PL provided one individual is able to fix the initial condition 
	$\vec b_0:=\lambda \vec j(0)+U'(\vec v^{\, 2}(0))\vec v(0)=\vec 0\,$. For a motion with vanishing initial speed one only 
	needs to impose $\vec j(0)=\vec 0\,$, irrespective of the choice of potential function $U$\,. 
	\item They may lead to a great variety of trajectories satisfying 2/3-PL according to the choice made for $U$. In the case of harmonic potential, elliptic trajectories are recovered, which are the best known case in which this law appears.
	\item The action variable $I_2=\frac{\lambda}{2\pi}\oint_\Gamma a\ dv\sim \int^{t_f}_{t_i}\lVert \vec j \rVert^{2}\ dt$ makes explicitly appear the mean squared jerk, and it is known that minimizing this function (maximizing smoothness) is an experimentally observed principle in motor control \cite{Flash2007}. Mechanics imposes that $I_2$ is constant but not necessarily minimal. However, provided $\lambda$ is a mass scale, $I_2$ has the dimension of the mechanical power. Minimizing $I_2$ during motion is therefore a way to minimize power expenditure. Figure \ref{fig} gives a schematic representation of $I_2$ in this case.
\end{itemize}

Beyond the confines of the 2/3-PL, humans exhibit various stereotyped behaviours, such as bell-shaped velocity profiles \cite{Atkeson2318} and adherence to Fitt's law \cite{Fitts54}. In the first case, the kinematics of horizontal reaching movements invariably shows a peak velocity in the middle of the trajectory, with a slight asymmetry depending on whether the movement is upward or downward \cite{Papa98}. These phenomena, which describe kinematic features of horizontal reaching movements and the compromise between speed and accuracy, are underpinned by the central nervous system's capacity to maintain consistency in the face of environmental variability. Our study suggests that theoretical constructs like action variables  may mirror such neural processes, serving as constants/constraints in the complex generation of human motion. The central nervous system, equipped with its intricate network of neurons and sensory receptors, plays a pivotal role in orchestrating movements. The brain, acting as the command centre, receives sensory feedback, processes it, and generates precise motor commands. It is conceivable that the brain, through its complex computations and feedback loops, sets dynamical principles and initial conditions that align with the physiological capabilities of the human body, ultimately giving rise to observed movement patterns, including the 2/3-PL.

An open challenge lies in identifying the neural correlates for these theoretical constructs. While certain brain regions responsible for sensory processing are well-understood, those involved in integrating multimodal sensory information to estimate physical forces like gravity remain less defined \cite{Stahn2023}. Moreover, the potential role of the brain and nervous system in shaping appropriate initial conditions and motor control strategies is  a crucial aspect of our proposal that remains to be elucidated. This not only affects the trajectories described by our mathematical models but also fundamentally influences the very ability of these models to mirror real human movement.

As a last comment we note that 2/3-PL and minimal jerk, that are closely linked within our framework, are not the only proposed models to describe voluntary motion. Let us consider Lagrangian (\ref{LFH2}) in which the position degree of freedom is actually an angular degree of freedom describing a circular motion in a plane, i.e., a circular motion 
restricted to a single constant axis of rotation. In this case the movement would tend to minimize $I_2\sim \int^{t_f}_{t_i}\dot\alpha^{\, 2}\ dt$, in which $\alpha$ is the angular acceleration. By virtue of Euler equation with a single rotation axis, $\dot\alpha\sim \dot M$ with $M$ the external torque applied on the considered body and $I_2\sim \int^{t_f}_{t_i}\dot M^{\, 2}\ dt$. 
This is a simplified version of the  minimal torque-change model \cite{Uno89}, suggesting that voluntary human motion minimizes the rate of change of total external torque. This last comment suggests that, beyond 2/3-PL, the integration of Lagrangian and Hamiltonian formulations of mechanics may fit within a general motor control model of voluntary motion encompassing several kinds of cost functions.

\end{document}